\documentclass[conference]{IEEEtran}
\IEEEoverridecommandlockouts
\usepackage{cite}
\usepackage{amsmath,amssymb,amsfonts}
\usepackage{algorithmic}
\usepackage{graphicx}
\usepackage{textcomp}
\usepackage{xcolor}
\def\BibTeX{{\rm B\kern-.05em{\sc i\kern-.025em b}\kern-.08em
    T\kern-.1667em\lower.7ex\hbox{E}\kern-.125emX}}
\begin{document}
	
	\title{Context-aware Attention-based Data Augmentation for POI Recommendation}
	
    \author{
        \IEEEauthorblockN{Yang Li}
        \IEEEauthorblockA{
        \textit{University of Queensland}\\
        Australia \\
        yang.li@uq.edu.au}
        \\
        \IEEEauthorblockN{Shazia Sadiq}
        \IEEEauthorblockA{\textit{University of Queensland} \\
        Australia \\
        shazia@itee.uq.edu.au}
        \and
        \IEEEauthorblockN{Yadan Luo}
        \IEEEauthorblockA{\textit{University of Queensland} \\
        Australia \\
        lyadanluol@gmail.com}
        \\
        \IEEEauthorblockN{Peng Cui}
        \IEEEauthorblockA{\textit{Tsinghua University} \\
        China \\
        cuip@tsinghua.edu.cn}
        \and
        \IEEEauthorblockN{Zheng Zhang}
        \IEEEauthorblockA{\textit{University of Queensland} \\
        Australia \\
        darrenzz219@gmail.com}
    }
    
	\maketitle
	
	\begin{abstract}
		With the rapid growth of location-based social networks (LBSNs), Point-Of-Interest (POI) recommendation has been broadly studied in this decade. Recently, the next POI recommendation, a natural extension of POI recommendation, has attracted much attention. It aims at suggesting the next POI to a user in spatial and temporal context, which is a practical yet challenging task in various applications. Existing approaches mainly model the spatial and temporal information, and memorize historical patterns through user's trajectories for recommendation. However, they suffer from the negative impact of missing and irregular check-in data, which significantly influences the model performance. In this paper, we propose an attention-based sequence-to-sequence generative model, namely POI-Augmentation Seq2Seq (PA-Seq2Seq), to address the sparsity of training set by making check-in records to be evenly-spaced. Specifically, the encoder summarises each check-in sequence and the decoder predicts the possible missing check-ins based on the encoded information. In order to learn time-aware correlation among user history, we employ local attention mechanism to help the decoder focus on a specific range of context information when predicting a certain missing check-in point. Extensive experiments have been conducted on two real-world check-in datasets, Gowalla and Brightkite, for performance and effectiveness evaluation. 
	\end{abstract}
	
	
	\section{Introduction}
	The huge explosion of location-based social networks (LBSNs) enables users to check in at real-world locations and share posts with others, which facilitates the development of location-aware services. Point-of-interest recommendation is one of the most extensively studied tasks among those services. Generally, users' check-in data contain textual, temporal and geographical information. This provides great opportunities for understanding user movement behaviours, such as users' short-term and long-term preferences, mobile patterns, based on which location-based service providers can finally tailor more accurate customer marketing strategies. Given this background, next POI recommendation is proposed targeting at recommending where a certain user is likely to visit in the next time period based on his check-in history.
	
	The recent work or studies on sequential data analysis and recommendation, latent representation models and the Markov chain models have been widely explored. Rendle et al.\cite{rendle2010factorizing} propose Factorizing Personalized Markov Chain (FPMC) for the next basket recommendation problem. After that, Cheng et al.\cite{cheng2013you} extend the FPMC model by embedding user movement constraint and setting the basket number to be 1 for next POI recommendation. Zhao et al.\cite{zhao2016stellar} introduce a time-aware model, namely STELLAR, which learns to score each successive POI via a ranking-based pairwise tensor factorization framework. In recent years, deep learning techniques have demonstrated magical power by delivering state-of-the-art performance in various research areas. Recurrent neural network (RNN) and its variants, such as Long-Short Term Memory (LSTM) and Gated Recurrent Unit (GRU), have become commonly implemented means for sequential data analysis tasks. Liu et al.\cite{liu2016predicting} develop a novel RNN-based model, called ST-RNN, which is designed to model temporal and spatial contexts via a set of discrete time and distance transition matrices. Zhao et al.\cite{zhao2018go} and Jarana et al.\cite{manotumruksa2018contextual} both add designed gates for spatial-temporal modelling inside RNN or RNN variants. Zhao et al.\cite{zhao2018go} implement a time gate and a distance gate in the standard LSTM cell to control the updating process of users' short-term and long-term interests, while, Jarana et al.\cite{manotumruksa2018contextual} aim at capturing the influences of time interval and distance factors by adding corresponding gates in GRU cell. Kong et al.\cite{kong2018hst} not only consider the impact of time and user movement but also introduce area-of-interest (AOI) into his hierarchical model.
	
	\begin{figure}[htb]
		\includegraphics[width=\linewidth]{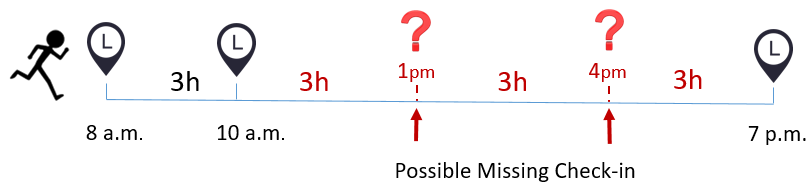}
		\caption{An illustration of data augmentation task in this paper. Assume that the user only checked in at 8 a.m., 10 a.m. and 7 p.m. in a day. We aim at imputing the possible missing check-in records at 1 p.m. and 4 p.m. to make the check-in sequence to be evenly-spaced with 3 hours.}
		\label{fig:augmentation-task}
	\end{figure}
	
	\begin{figure}[htb]
		\includegraphics[width=\linewidth]{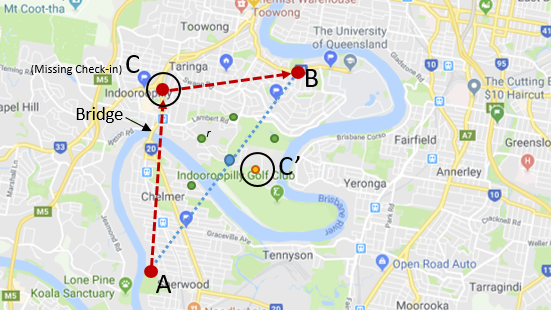}
		\caption{An illustration of an failure example of linear interpolation method, where $C$ denotes a POI that the user has visited after checked in POI $A$ and before visiting POI $B$. $C^{'}$ indicates the estimated check-in point which is generated via linear interpolation.}
		\label{fig:missing_checkin}
	\end{figure}
	
	Nevertheless, the existing methods mentioned above are affected by the negative influence of data sparsity issue to some extent. Users' check-in data collected from LBSNs is much more sparse than the general item-rating dataset. In movie or shopping rating data, users are more likely to give comments and ratings on the items or movies they have watched or purchased. However, it is unlikely for a user to check in every time when he visits a POI. As a result, the datasets for next POI recommendation naturally lose many possible check-in records between each consecutive check-in pairs. Suffering from incomplete user check-in trajectories, the existing models are hard to learn accurate user transition patterns. For instance, as shown in Figure \ref{fig:missing_checkin}, only the locations $A$ and $B$ were checked in and the possible check-in $C$ is missing. It will be hard for RNN to learn the $A\rightarrow C$ and $C\rightarrow B$ transition patterns. Therefore, it is reasonable that adding possible missing check-in data into the existing dataset can help to improve the next POI recommendation model performance and stability.
	
	To the best of our knowledge, this is the first study on data augmentation for next POI recommendation task. As illustrated in Figure \ref{fig:augmentation-task}, our task is to insert possible missing check-in data into the training set making it to be evenly-spaced in terms of time interval. Finally, helping the next POI recommendation model better understand user preference and behaviour patterns. Traditionally, a common way for data augmentation is linear interpolation, which, in this task, calculates the point $p$'s geographical position on straight line between two observed POIs and the the nearest POI $l$ of $p$ is selected as the predicted missing check-in. However, this method is unreliable, since the user trajectory path is more likely to be a curve rather than straight line due to the geographical constraint and temporal preference. Therefore, it could be easy to imagine that the some points imputed by linear interpolation are far away from the actual check-in locations, which misleads learning models. To capture and model the actual user trajectories in real-world, we propose to address the drawbacks described above through an encoder-decoder model, namely POI-Augmentation Sequence-to-Sequence model (PA-Seq2Seq). More specifically, as illustrated in Figure \ref{fig:overview}, the model maps the check-in sequence masked with missing check-in tokens denoted as $mc$ in Figure \ref{fig:overview} to context vectors by a multi-layer encoder and outputs missing check-ins with a multi-layer decoder conditioned on encoded context information. 
	To better utilise the context based on the current target check-in,  we choose to adopt local attention mechanism that can guide the decoder to pay attention to a specific range of contextual information for better missing record imputation performance. Moreover, we propose a multi-stage training strategy including maximum log likelihood, randomly masking to help our model alleviate over-fitting problem and coverage faster.
	
	Our main contributions are summarized as follows:
	\begin{itemize}
		\item To the best of our knowledge, it is the first work to investigate the advanced augmentation techniques for next POI recommendation task.
		\item An attention-based sequence-to-sequence model is developed for check-in data generation with multiple training stages, which guarantees the stability and performance of our model.
		\item Extensive experiments have been conducted, which demonstrates that the proposed model outperform linear interpolation methods in both imputation accuracy and augmentation effectiveness.
	\end{itemize}
	
	The remaining parts of this paper are organized as follows. In Section 2, we discuss the advances of next POI recommendation, data augmentation as well as sequence-to-sequence models. Section 3 describes the details of the proposed model, PA-Seq2Seq, which is followed by experimental results for proving the performance and effectiveness of this model in Section 4. Finally, we give a conclusion and represent our future work in Section 5 and Section 6 respectively.
	
	\section{Related Work}
	In this section, we first review a number of methods for next POI recommendation, which can be categorised as matrix factorization methods, the Markov chain methods and RNN-based methods. Then, we review the advances of sequential data augmentation techniques and finally describe the development of sequence-to-sequence model.
	
	\subsection{Next POI Recommendation}
	Next POI recommendation is recently proposed and has received great research interests. It is an extension of conventional POI recommendation, which pays more attention to sequential influence as well as spatial and temporal context compared with the traditional one. Cheng et al.\cite{cheng2013you} recommend successive POI by decomposing tensors which are constructed utilising the first-order Markov chain with geographical distance constraints. He et al. \cite{he2016inferring} argue that temporal and categorical information plays an important role in next POI recommendation. Thus, they develop a tensor-based latent model that studies user's behaviour patterns via personalized Markov chain and train the model through Bayesian Personalised Ranking (BPR) \cite{rendle2009bpr}. Feng et al. \cite{feng2015personalized} employ embedding technique to jointly consider check-in sequence transition pattern and user's preferences in recommendation. Furthermore, they extend the proposed model PRME to PRME-G whose ranking metrics are incorporated with geographical influences when predicting the next POI. Xie et al. \cite{xie2016learning} adopt bipartite graph in their graph-based embedding learning model called GE for next location recommendation task. The success of recurrent neural network (RNN) in various areas attracts many researcher to use RNN and its variants for this problem. Liu et al. \cite{liu2016predicting} propose a RNN-based model ST-RNN. They replace the standard weight matrix in RNN cell with time-specific and distance-specific transition matrices, which can help to capture temporal and spatial influences. The original RNN can easily cause gradient vanish and gradient explosion during the training process. Long-short-term memory (LSTM) \cite{Hochreiter:1997:LSM:1246443.1246450}, a popular variant of RNN, is proposed to address the issues mentioned above. It has three different gates: input gate, output gate and forget gate, which largely alleviate the gradient vanish and explosion problems. Several recent work focusing on modifying gating mechanisms inside LSTM cell to better learn temporal and spatial information from check-in sequential data. Zhu et al. \cite{zhu2017next} modify the inner structure of standard LSTM cell by adding time-specific gates for time-aware next item recommendation. Kong et al. \cite{kong2018hst} design the ST-LSTM whose three standard LSTM gates are enriched by spatial-temporal specific weight matrices. They believe the influence of distance and time can offer learning guidance for the proposed model in such manner. They further extend ST-LSTM by employing hierarchical architecture. More specifically, incorporating ST-LSTM in encoder-decoder structure for modelling contextual visiting history, which boosts model performance. Inspired by coupled LSTM architecture \cite{greff2017lstm}, Zhao et al. propose ST-CLSTM which has carefully design coupled time and distance gates for capturing spatial and temporal contextual information. The model achieves state-of-the-art performance in next POI recommendation problem.
	
	\subsection{Time-series Data Augmentation}
	Data augmentation includes a set of methods that introduce unobserved data or latent variables via sampling algorithms or other generative models. Such techniques are frequently employed in time series data, since imputation of missing values can be of benefit in various types of time-series data-driven applications, such as health care \cite{che2018recurrent}, financial market, etc. Several statistical methods have been studied for years, ARIMA \cite{ansley1984estimation} and state space model \cite{lee1994state} eliminate non-stationarity in the time-series dataset for missing data imputation. Recent works utilise deep learning techniques, e.g., recurrent neural networks for data augmentation task. GRU-D is proposed by Che et al. \cite{che2018recurrent} that predicts the missing value of health data by combining the last observed value and global mean. In this paper, we investigate the solution on augmenting POI sequence data for better recommendation results.
	
	\subsection{Sequence-to-sequence Models and Attention Mechanism}
	The sequence-to-sequence models have achieved satisfactory results in various areas, such as machine translation \cite{luong2015effective,wu2016google,vaswani2018tensor2tensor,shen2015minimum}, dialogue systems \cite{duvsek2016sequence}, text summarisation \cite{nallapati2016abstractive,see2017get}, headline generation \cite{rush2015neural,nallapati2016abstractive}, speech to text conversion \cite{graves2013speech, chorowski2015attention} and image captioning \cite{xu2015show,vinyals2015show,karpathy2015deep}. This general framework comprises two main components: encoder and decoder. The encoder reads the sequence of data from input and the decoder uses the outputs passed from the encoder to produce a sequence of tokens used as final outputs. Sutskever et al. \cite{sutskever2014sequence} and Luong et al. \cite{luong2014addressing} both present a general sequence-to-sequence model which consists of multilayered LSTM in both encoder and decoder for machine translation task. However, the tradition encoder-decoder model does not work well on long sequences. This is because the encoder compresses all the information of a source sequence into a fixed-length vector. Cho et al. \cite{cho2014properties} show that the performance of a basic encoder-decoder deteriorates quickly when the input length increases. To address such an issue, a number of studies employ attention mechanism to extend the basic Seq2Seq model. Bahdanau et al. \cite{bahdanau2014neural} propose an attention mechanism that can automatically search for the sentence parts which are related to the target translating word. Luong et al. \cite{luong2015effective} combine the hard attention and soft attention mechanism for translation. Specifically, the attention module will calculate an alignment centre and compute the attention score according to a range of source hidden states around the this centre.  In this way, the computation consumption is largely reduced as the model only needs to calculate a small range of hidden states for attention weights. 
	
	\section{Proposed Method}
	In this section, we first describe the key definitions used in this paper and formulate our problem, and then present the details of our proposed model POI Augmentation Seq2Seq. 
	
	\subsection{Problem Formulation}
	Let $S=\{s_1, s_2, ..., s_m\}$ be the set of training data, where each entry $s$ denotes a sequence of check-in data $s = \{c_1,c_2,...,c_n\}$, and $c_t$ indicates a user-place-time tuple $(u, l, t)$ for simplicity. $\Delta t$ and $\Delta d$ denote the time interval and distance between two consecutive check-ins $c_t$ and $c_{t+1}$ respectively. Our target is to address the sparsity issue of training data $S$ by imputing each possible missing check-in record denoted as $mc$ to enrich the training set and make each check-in sequence $s$ to be evenly-spaced. Consequently, the performance of POI recommendation will be boosted to a large extend with the augmentation by our proposed method.
	
	\begin{figure}[ht]
		\includegraphics[width=\linewidth]{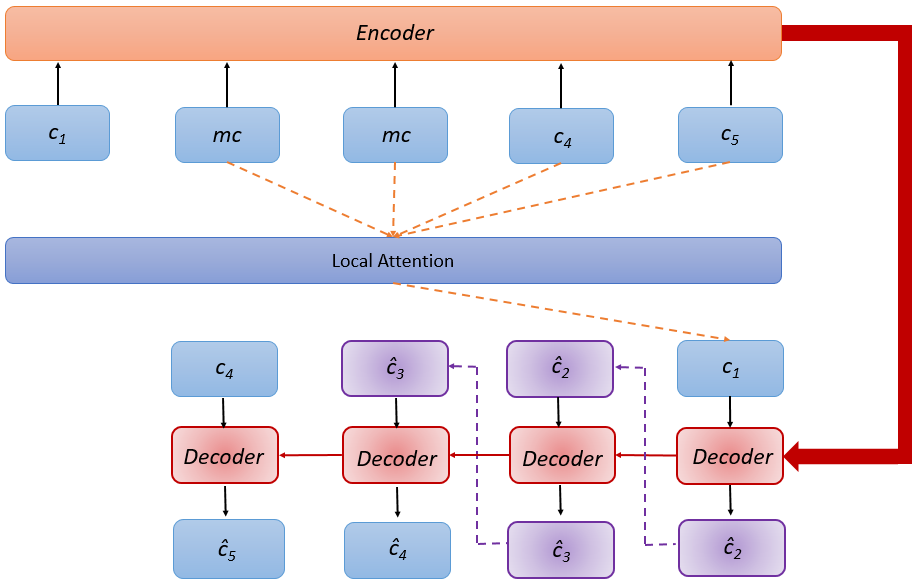}
		\caption{The illustration of general architecture of the proposed PA-Seq2Seq. It is consisted of three major parts: encoder network, decoder network and attention network which will be further introduced in Figure \ref{fig:decoder}. The $mc$ with blue box denotes a missing check-in. It is treated in the same way as the normal check-ins (e.g., $c_1$ and $c_5$), when they are fed into the encoder. The encoder compresses the check-in sequence into context vector and passes it to the decoder via red arrow line. Meanwhile, the decoder predicts missing check-in based on the context.}
		\label{fig:overview}
	\end{figure}
	
	\subsection{Attention-based POI Augmentation Sequence-to-sequence Model}
	The missing check-in imputation task requires the model to consider both past and future check-in contextual information, and hence, an encoder-decoder model combined with attention mechanism is adopted for this task. The general framework architecture of the proposed POI-Augmentation Sequence-to-Sequence model (PA-Seq2Seq) is shown in Figure \ref{fig:overview}. It has three components, encoder network, decoder network and attention network.
	
	\begin{figure}[ht]
		\includegraphics[width=\linewidth]{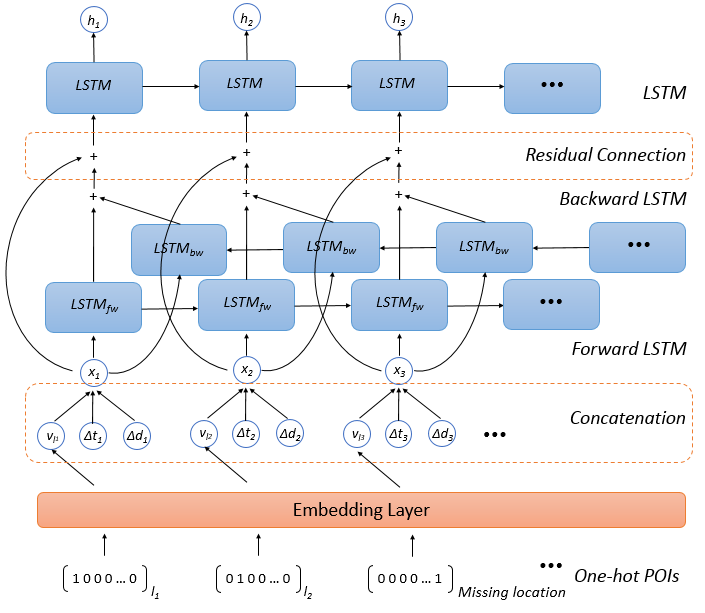}
		\caption{An illustration of detailed encoder structure. The embedding layer converts one-hot POI encoding to a $d$-dimension vector $v$. The concatenation operation fuses $v$, $\Vec{\Delta t}$ and $\Vec{}\Delta d$ into one vector $x$. The whole model comprises of a bi-directional LSTM and stacked with a uni-directional LSTM. They are connected using residual connection mechanism for more effective gradient flow in training time.}
		\label{fig:encoder}
	\end{figure}
	\begin{figure*}[ht]
		\centering
		\includegraphics[width=\linewidth]{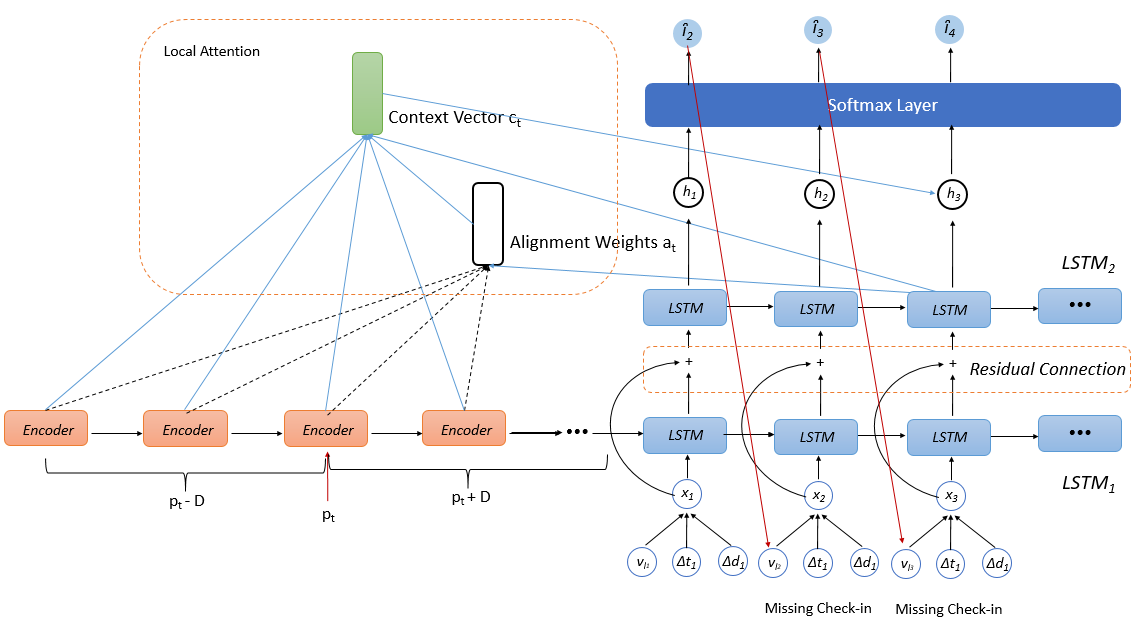}
		\caption{An illustration of the proposed decoder structure which employs residual connection and local attention mechanism. The right part shows that the decoder comprises of a two-layer stacked LSTM connected with residual connection. The second and third check-ins are missing on the figure, therefore, the outputs $\hat{l}_2$ and $\hat{l}_3$ are feed into the model shown as the red arrow line. In the left part, $p_t$ indicates the align centre when predicting $\hat{l}_3$. The attention alignment weights $a_t$ are calculated based on the hidden states from $p_t -D$ to $p_t + D$ (linked with black dashed arrow line) and the current target decoding hidden state $h_3$.The context vector $c_t$ is then computed according to $a_t$ and hidden states from window $[p_t-D,p_t+D]$. The attention-aware hidden states are denoted as $h_1,h_2$ and $h_3$. Finally, the decoder outputs the final prediction through a softmax layer colored with dark blue, which converts the hidden state into a check-in probability distribution.}
		\label{fig:decoder}
	\end{figure*}
	\subsection{Stacked Encoder for Check-in Sequence Annotation}
	The missing check-in imputation task requires our model to generate possible missing check-in tokens based on the given context. Therefore, a stacked-LSTM is proposed to be used as encoder. The Figure \ref{fig:encoder} demonstrates this encoder structure. The bi-directional LSTM is the first layer consists of forward and backward LSTMs. It is stacked by a uni-directional LSTM. When encoding a check-in sequence at each timestep $t$, a concatenated vector $x_t$, which contains POI representative vector $v_{l_t}$, time-specific vector $\Vec{\Delta t_t}$ and distance-specific vector $\Vec{\Delta d_t}$, is fed into the model (i.e., $x_t = [v_{l_t},\Vec{\Delta t_t}, \Vec{\Delta d_t]^\top}$). The forward LSTM reads the fed data from the first one $c_1$ to the last one $c_n$, while the backward one reads in reverse order, i.e., $\{c_n, c_{n-1}...c_1\}$. The hidden state $h_t$ and cell state denoted as $cs$ produced from both LSTM layers are concatenated to $h_{BiLSTM}^t$ at each timestep $t$, it can be calculated in the following equation:
	\begin{equation}\label{first_layer}
	\begin{aligned}
	h_t^{fw} &= LSTM_{fw}(h^{fw}_{t-1}, cs^{fw}_{t-1}, x_t)\\
	h_t^{bw} &= LSTM_{bw}(h^{bw}_{t-1}, cs^{bw}_{t-1},x_t)\\
	h^{bilstm}_t &= [h_t^{fw}, h_t^{bw}]^\top\\
	cs^{bilstm}_t &= [cs_t^{fw}, cs_t^{bw}]^\top
	\end{aligned}
	\end{equation} 
	Inspired by Wu et al. \cite{wu2016google}, instead of using direct connection, we employ residual connection structure between two stacked layers for information propagation. It empirically improves the gradient flow of the model, and thus is helpful for training each layer more effectively, which is expected to receive better training result than the directly-connected one. Specifically, in a common stacked LSTM model, the inputs for the second layer are hidden states output from first layer. Therefore, both layers' hidden states can be calculated as:
	\begin{equation}
	\begin{aligned}
	cs^1_t, h^1_t &= BiLSTM(cs^1_{t-1}, h^1_{t-1}, x^0_t) \\
	x^1_t &= h^1_t \\
	cs^2_t, h^2_t &= LSTM(cs^2_{t-1}, h^2_{t-1}, x^1_t)
	\end{aligned}
	\end{equation}
	However, with residual connection, as shown in Figure \ref{fig:encoder}, the inputs for the next layer are changed to the concatenation of the inputs and hidden states from the previous layer. More concretely, at timestep $t$, for the proposed two-layer stacked LSTM with residual connection, the formula is changed to:
	\begin{equation}
	\begin{aligned}
	cs^1_t, h^1_t &= BiLSTM(cs^1_{t-1}, h^1_{t-1}, x^0_t) \\
	x^1_t &= h^1_t + x^0_t\\
	cs^2_t, h^2_t &= LSTM(cs^2_{t-1}, h^2_{t-1}, x^1_t)
	\end{aligned}
	\end{equation}
	\subsection{Attention-based Decoder for Check-in Imputation}
	The decoder works to conditionally impute the possible missing check-in tokens. If the previous check-in is a missing one, the predicted one will be fed into the decoder. In order to make the missing token imputation more accurate. it is necessary to consider a set the past and future check-in records around the current check-in token. Thus, the decoder is integrated with attention mechanism, which is similar to \cite{luong2015effective}. When predicting a missing check-in token, the last check-in plays the most important role in prediction. Therefore, we set the alignment pointer $p_t$ targets at the last check-in timestep $t$, i.e., $pt=t$. To capture user's temporal preference at $t$, we set the window size $D$ to be 10, which means when predicting the missing check-in at time step $t$, the model takes the information of check-ins $[c_{t-D},...,c_{t+D}]$ into account by putting a Gaussian distribution centred around the last check-in $c_t$. Therefore, the alignment vector $a_t$ can be computed as:
	\begin{equation}
	\begin{aligned}
	score(h_t,\overline{h_s}) &= h_t^\top W_a\overline{h_s}\\
	a_t(s)&=align(h_t,\overline{h_s})exp(-\frac{(s-p_t)^{2}}{2\sigma^2})\\
	&=\frac{exp(score(h_t,\overline{h_s}))}{a}
	\end{aligned}
	\end{equation}
	where $h_t$ denotes the current check-in's hidden state and $\overline{h_s}$ denote the source states in the design attention window. The general attention score function is adopted in this paper.
	\subsection{Zoneout and Mask Training Strategy}
	To make our model more robust, we adopt zoneout regularisation approach \cite{krueger2016zoneout} during the training stage, which randomly preserves the hidden states. Under this POI check-in sequence context, the zoneout mechanism actually randomly removes a part of check-in information. This way facilitates our model to capture the unobserved check-ins in a check-in sequence through backpropagation. It also makes the gradient flow to earlier timesteps more effectively. As a result, the model is able to study the check-in distribution better and alleviate over-fitting problem. Apart from this mechanism, a mask training strategy is also employed to guide the model learn to impute missing check-ins. More concretely, when training the model, a set of observed check-ins are randomly selected and replaced by missing check-in tokens (i.e., the index of missing check-in token in the one-hot table is the number of total check-ins). As the example shown in Figure \ref{fig:decoder}, the model needs to predict the masked check-in token $mc_t$ conditioned on the masked context $m(s)$ from the encoder and the information of check-in sequence $\{c_1, c_2,..., c_{t-1}\}$. 
	
	\section{Experiment}
	
	\begin{table*}[htb]
		\caption{Performance of HR@1, HR@5 and HR@10 between various next POI recommendation methods trained on original Gowalla dataset and the ones augmented by linear interpolation and PA-Seq2Seq respectively.}
		\begin{tabular}{|c|c|c|c|c|c|c|c|c|c|c|c|c|}
			\hline
			Methods  & \multicolumn{3}{|c|}{Original}        &\multicolumn{3}{|c|}{Linear Interpolation (POP)}        & \multicolumn{3}{|c|}{Linear Interpolation (NN)}        &\multicolumn{3}{|c|}{\textbf{PA-Seq2Seq}}        \\
			\hline
			& HR@1    & HR@5 & HR@10 & HR@1 & HR@5 & HR@10 & HR@1   & HR@5 & HR@10 & HR@1   & HR@5             & HR@10 \\
			\hline
			FPMC-LR  & 0.029    & 0.052 & 0.085  & 0.030 & 0.053 & 0.087  & 0.033   & 0.057 & 0.092  & \textbf{0.035}   & \textbf{0.060} & \textbf{0.097}  \\
			PRME-G  & 0.034    & 0.065 & 0.087  & 0.038 & 0.070 & 0.091  & 0.042   & 0.081 & 0.098  & \textbf{0.042}   & \textbf{0.091} & \textbf{0.122}  \\
			RNN      & 0.064    & 0.129 & 0.170  & 0.066 & 0.133 & 0.173  & 0.066   & 0.148 & 0.191  & \textbf{0.073}   & \textbf{0.155}             & \textbf{0.200}  \\
			LSTM     & 0.073    & 0.151 & 0.191  & 0.079 & 0.158 & 0.198  & 0.084   & 0.164 & 0.205  & \textbf{0.089}   & \textbf{0.171}             & \textbf{0.215}  \\
			ST-CLSTM & 0.085    & 0.147 & 0.179  & 0.090 & 0.162 & 0.195  & 0.091   & 0.163 & 0.196  & \textbf{0.095}   & \textbf{0.172}             & \textbf{0.207}  \\
			
			\hline
		\end{tabular}
		\label{table:gowalla_eval}
	\end{table*}
	
	\begin{table*}[htbp]
		\centering
		\caption{Performance of various next POI recommendation approaches evaluated by HR@1, HR@5 and HR@10, which are trained on original Brightkite dataset and augmented ones respectively.}
		\begin{tabular}{|c|c|c|c|c|c|c|c|c|c|c|c|c|}
			\hline
			Methods  & \multicolumn{3}{|c|}{Original}        &\multicolumn{3}{|c|}{Linear Interpolation (POP)}        & \multicolumn{3}{|c|}{Linear Interpolation (NN)}        &\multicolumn{3}{|c|}{\textbf{PA-Seq2Seq}}        \\
			\hline
			& HR@1    & HR@5 & HR@10 & HR@1 & HR@5 & HR@10 & HR@1   & HR@5 & HR@10 & HR@1   & HR@5 & HR@10 \\
			\hline
			FPMC-LR  & 0.163    & 0.247 & 0.316  & 0.168 & 0.255 & 0.336  & 0.187   & 0.284 & 0.354  & \textbf{0.195}   & \textbf{0.296} & \textbf{0.372}  \\
			PRME-G  & 0.197    & 0.299 & 0.349  & 0.221 & 0.312 & 0.352  & 0.235   & 0.257 & 0.362  & \textbf{0.245}   & \textbf{0.321} & \textbf{0.388}  \\
			RNN      & 0.408    & 0.468 & 0.489  & 0.413 & 0.480 & 0.499  & 0.423   & 0.465 & 0.502  & \textbf{0.430}   & \textbf{0.495} & \textbf{0.510}  \\
			LSTM     & 0.356    & 0.445 & 0.483  & 0.364 & 0.454 & 0.482  & 0.379   & 0.460 & 0.483  & \textbf{0.396}   & \textbf{0.464} & \textbf{0.488}  \\
			ST-CLSTM & 0.446    & 0.496 & 0.522  & 0.456 & 0.495 & 0.517  & 0.450   & 0.499 & 0.523  & \textbf{0.457}   & \textbf{0.512} & \textbf{0.543}  \\
			\hline	
		\end{tabular}
		\label{table:brightkite_eval}
	\end{table*}
	
	We conduct extensive experiments to prove the effectiveness of our model. In this section, we first briefly introduce the datasets used in the experiment and then describe the details of our method implementation. 
	
	\subsection{Dataset Description}
	We train and evaluate our model on two public large-scale extremely sparse LBSN check-in datasets, Gowalla and Brightkite, with density 0.012\% and 0.209\% respectively.
	
	\textbf{Gowalla Dataset}:
	It is a location-based social network mobile application that was launched in 2007 and eventually shut down in 2012 due to the acquisition by Facebook. Users can check in a named POI and share with their friends through this app. In this paper, a complete snapshot of Gowalla check-in data is used. The dataset contains 18,737 POIs, 32,510 users with 1,278,274 check-in records over the period of February 2009 to October 2010.
	
	\textbf{Brightkite Dataset}:
	This dataset was collected during the period from April 2008 to October 2010. Brightkite enables users check in via text messaging or application. Users can also connect with their friends and meet new friends through viewing the places they have previously visited. There are 4,747,288 check-in records containing 51,406 POIs and 772,967 users in this dataset.
	
	\subsection{Experiment Details}
	The training process for PA-Seq2Seq has three stages. Initially, a uni-directional LSTM and bi-directional LSTM are trained via maximum likelihood estimation (MLE) respectively. The weights of pretrained models will be used as encoder and decoder initialization. In the second training stage, the PA-Seq2Seq is also pretrained via MLE. Finally, in the third training stage, the model is trained through mask training strategy, which has already been described in Section 3.E. Note that the mask percentage increases iteratively from 10\% at first epoch to 50\% at 50th epoch. This three-stage training can keep the training process to be stable and the model coverage faster. We set the the dimensionality of POI embeddings to be $16$. We use Adam optimizer \cite{kingma2014adam} for gradient decent and set the learning rate to be 0.008 in the experiment.
	
	\subsection{Baseline Methods}
	To the best of our knowledge, missing check-in imputation is the first time to be used in the next POI recommendation. Therefore, we compare the proposed model with two types of basic linear interpolation methods both assuming users tend to go through the two POIs between $l_1$ and $l_2$ along the shortest path. Assume there are two observed check-ins $c_1$ and $c_2$ associated with their coordinates $\{lat_1, lng_1\}$ and $\{lat_2, lng_2\}$. The point $p_1$
	One way is to select the nearest neighbour according to the calculated point $p$, while the other one chooses the most popular POI around $p$.
	
	\subsection{Next POI Recommendation Methods}
	There are a number of next POI recommendation 
	\begin{itemize}
		
		\item FPMC-LR\cite{cheng2013you}: The model constructs transition probability matrices via personalised Markov chain approach. It factorizes the transition tensor of each user under the user movement constraints around a localized region to calculate the transition probability of selected POIs when predicting next location.
		
		\item PRME-G\cite{feng2015personalized}: The method projects users and POIs into the latent spaces to calculate feature similarity be between users and POIs for recommendation.
		
		
		\item LSTM\cite{Hochreiter:1997:LSM:1246443.1246450}: A variant of recurrent neural network, which has a memory cell, an input gate, an output gate and a forget gate. It largely helps to solve gradient vanishing problem, and enables long-term dependency learning.
		
		\item RNN\cite{zhang2014sequential}: This method uses standard recurrent neural network to learn temporal dependencies in user's behavioural sequential information.
		
		\item ST-CLSTM\cite{zhao2018go}: The state-of-art approach that integrates coupled input and forget gates to capture spatio-temporal interval information between check-ins.
		
	\end{itemize}
		
	\subsection{Evaluation Methods}
	We first separate the dataset according to each record's user ID. Then, each user's check-in records are ordered by timestamp. The first 80\% of check-ins are used as training set and the remaining ones are used for testing. In the training set, the last 10\% of check-in data are used for validation to help tune hyper-parameters of the model. 
	
	We use Hit Ratio at $K$ (HR$@K$) metric, which is widely adopted in \cite{yin2013lcars,wang2015geo,yin2016adapting,yin2015joint} to evaluate our effectiveness of our method. The formula of HR$@k$ is given as follows:
	\begin{equation}
	\begin{aligned}
	HR@k=\frac{\#hit@k}{|test|}
	\end{aligned}
	\end{equation}
	where $\#hit@k$ indicates the number of times that the ground truth $l$ appears in the top-k ranking result produced by the model. $\#hit$ is added by $1$ if $l$ is in the result list, otherwise, keeps the same value. $|test|$ denotes the number of test cases.
	\subsection{Results Analysis}
	The performance comparison on the two datasets is evaluated by HR$@1$, HR$@5$ and HR$@10$ which are illustrated in Table \ref{table:gowalla_eval} and Table \ref{table:brightkite_eval}. The the training set augmented by the proposed model PA-Seq2Seq gives significant performance boost to all the next POI recommendation models achieve better accuracy performance. The HR$@1$, HR$@5$ and HR$@10$ of the state-of-the-art method ST-CLSTM increases HR$@10$ from 0.179 to 0.207 on Gowalla dataset and from 0.522 to 0.543 on Brightkite dataset. This proves that our model is able to extract user's latent visiting distribution from the both dataset. It can be observed that, linear interpolation methods can also improve all the existing methods' performance, but the effectiveness still lower than PA-Seq2Seq by relatively around 4\%. It is likely because the linear interpolation methods ignore the temporal and spatial context of the dataset.
	
	\subsection{Visualisation}
\begin{figure}[ht]
		\centering
		\includegraphics[width=0.48\linewidth, height=0.4\linewidth]{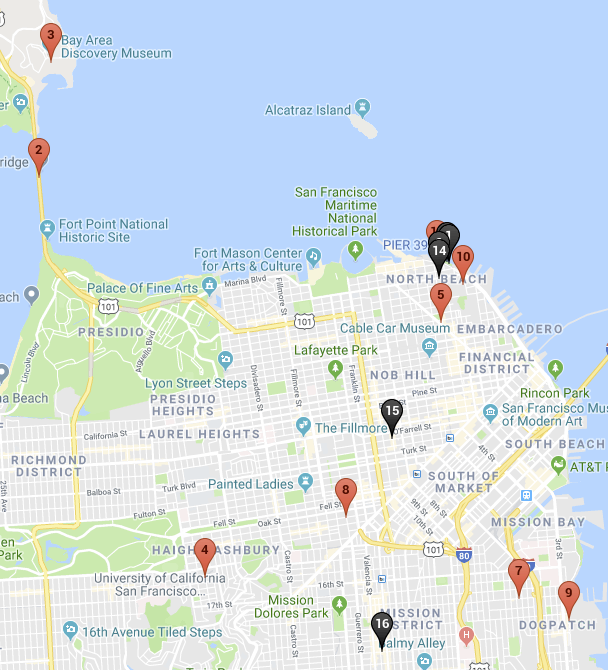}
		\includegraphics[width=0.48\linewidth, height=0.4\linewidth]{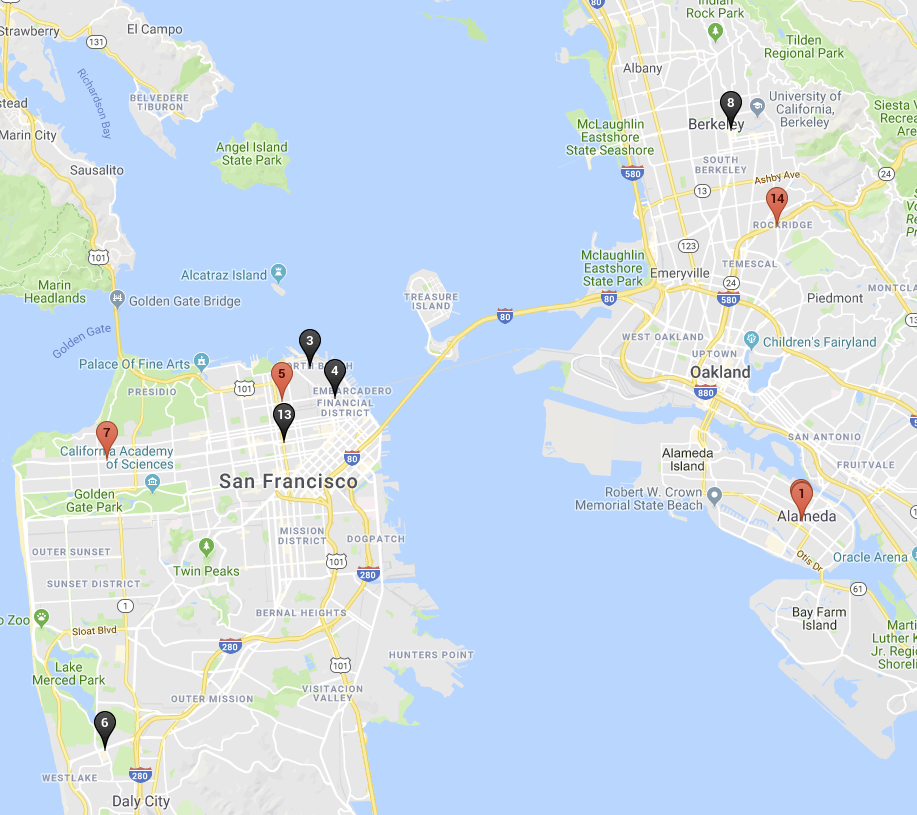}
		\caption{Two examples from Gowalla dataset before and after augmentation.}
		\label{fig:gowalla_visual}
	\end{figure}

\begin{figure}[ht]
		\centering
		\includegraphics[width=0.48\linewidth, height=0.45\linewidth]{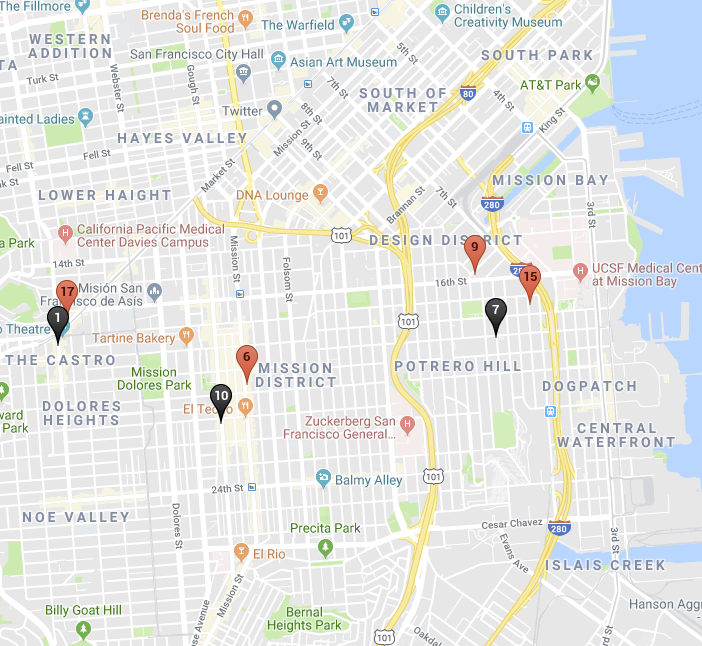}
		\includegraphics[width=0.48\linewidth, height=0.45\linewidth]{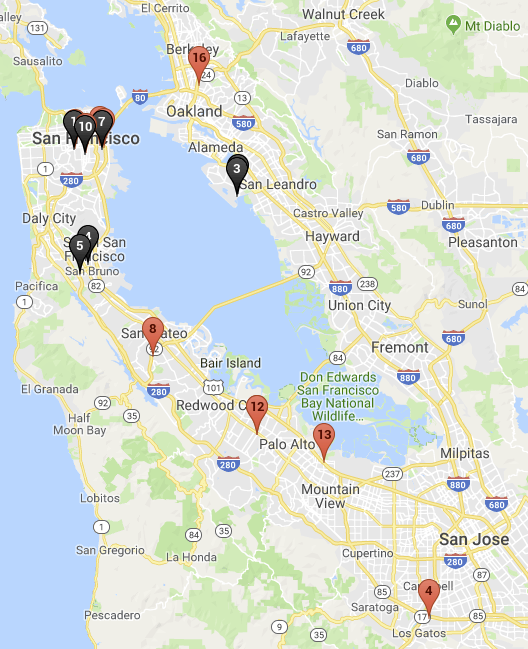}
		\caption{Two examples from Brightkite dataset before and after augmentation.}
		\label{fig:brightkite_visual}
	\end{figure}

	As illustrated in Figure \ref{fig:gowalla_visual} and Figure \ref{fig:brightkite_visual}, the black icons are the original check-in data, while the red icons are augmented data imputed by our model. The numbers on the icons are the order of a check-in sequence. The results have demonstrated that the model successfully model user's transition pattern to some extent, though there are some augmented check-ins seem not so reasonable on the map. The main reason is that both datasets are extremely sparse, it is common to see that the distance of two consecutive check-ins are larger than 10km, which is extremely hard for our model to correctly make prediction.
	\section{Conclusion}
	In this paper, an attention-based sequence-to-sequence model incorporating spatial and temporal information, named PA-Seq2Seq, is proposed to augment the check-in datasets for next POI recommendation. Our model learns missing check-in using an encoder-decoder model. The experiment results show that our model is able to boost the next POI recommendation model performance significantly. Additionally, different from linear interpolation, the model can also be applied in the next POI recommendation task directly, as it has learned the visiting distribution through training.
	
	\section{Future work}
	As mentioned in Section 4.G, two consecutive check-ins may be posted in two different cities by the user, it is reasonable to consider the geographical factor in the future. Besides, although making check-in data to be evenly-space by a manually-designed time length has received promising result, we can also investigate how to model a trajectory automatically according to the given-length time slot, starting place and end place, i.e., the number of augmented check-ins and check-in time can vary. Moreover, since our model not only can be used for data augmentation task, but also is capable to make prediction in next POI recommendation task. We would like to further evaluate the performance of our model for next POI problem. Apart from this, since the model for data augmentation task requires the starting and end point as well as time slots, which is very similar to the trip recommendation task that aims at recommending a trip plan based on the given departure place and destination place under the time limit. Therefore, enlightened by the successful application of generative adversarial networks in NLP, such as SeqGAN \cite{yu2017seqgan} and MaskGAN \cite{fedus2018maskgan}, we would like to adopt adversarial training strategy in our model for trip recommendation through learning from the real travelling historical data in the future. \cite{rtree,beckmann1990r,sellisRF87}
	\bibliographystyle{IEEEtran}
	\bibliography{bibliography}
\end{document}